\documentclass{revtex4}

\usepackage{amsmath}
\usepackage{amssymb}
\usepackage{graphicx}

\begin{document}

\title{Gravitational Model of the String}
\author{Vladimir Dzhunushaliev}
\email{dzhun@hotmail.kg}
\affiliation{Dept. Phys. and Microel. Engineer., Kyrgyz-Russian
Slavic University, Bishkek, Kievskaya Str. 44, 720021, Kyrgyz
Republic}

\date{\today} 

\begin{abstract}
It is shown that an infinite gravitational flux tube solution in 5D Kaluza-Klein gravity  with the cross section in the Planck region after $5D \rightarrow 4D$ reduction and isometrical embedding in a Minkowski  spacetime can be considered as a moving infinite string-like object. Such object has a flux of the electric and magnetic fields. The 4D gravitational waves on the tube are considered. 
\end{abstract}

\maketitle

\section{Introduction}

The string in string theory is an 1-dimensional structureless object. 
Historically the concept of string in the modern physical context has 
arisen in quantum chromodynamics (QCD). In QCD the confinement problem 
is connected with a hypothesized flux tube filled with the longitudinal 
color electric field and stretched between quark and antiquark. If we tend 
the cross section of the flux tube to zero we obtain a string which can 
give us some information about quark confinement. 
The problems with such string approximation for confinement are well known. 
These problems have resulted to the fact that the string is an 1-dimensional 
fundamental structureless object in the modern string theory. Nevertheless 
there is an open question : does have any field theory string-like 
solutions ? If yes then it is necessary to compare the classical and 
quantum properties such string-like solutions with real strings in string 
theory. Evidently the field theories like gauge theories will have the 
same problems with the quantization of string-like solutions as the quantization of string in QCD. 
\par
In this notice we will show that in the 5D Kaluza-Klein gravity there is a solution which  can be considered as a string-like object in some external embedding spacetime. The idea is  that any pseudo-Riemannian spacetime can be locally and isometrically embedded in an  Minkowski spacetime $\mathbb M$ \cite{Friedman}. The local topology of the discussed  solution is $\mathbb R_1 \times \mathbb R_2 \times \mathbb S^1 \times \mathbb S^2$ 
where the time is spanned on $\mathbb R_1$, the 5$^{th}$ coordinate on 1D sphere 
$\mathbb S^1$, the longitudinal coordinate on $\mathbb R_2$ and the polar angles 
$\theta, \varphi$ on 2D sphere $\mathbb S^2$. The factor $\mathbb R_2 \times \mathbb S^2$ describes a tube. If the cross section of the tube and $S^1$ can be chosen arbitrary small  then after embedding in the above mentioned Minkowski spacetime $\mathbb M$ we will have a  string-like object moving in $\mathbb M$. The most significant difference with the ordinary string is  that $\mathbb M$ can have more than one time. 

\section{5D flux tube}
\label{sec:5DFluxTube}

At first we will present the regular 5D wormhole-like flux tube metric 
which is the solution of the 5D vacuum Einstein's 
equations \cite{dzsin1} 
\begin{equation}
\begin{split}
	ds^2 & = \cosh^2 \left( \frac{r}{r_0} \right) dt^2 - 
		dr^{2} - r^2_0 \left( d\theta ^{2} +
	\sin ^{2}\theta  d\varphi ^2 \right ) 
\\
	& -	\left [d\chi + 
	\omega(r) dt + 
	Q \cos \theta d\varphi \right ]^2 , 
\\ 
	\omega(r) & = \sqrt{2} \sinh \left( \frac{r}{r_0} \right),
	\quad 
	Q = \sqrt{2} r_0
\label{sec1-10}
\end{split}
\end{equation}
where $0 \leq \chi \leq 2 \pi r_0$ is the 5$^{th}$ extra coordinate; 
$\theta ,\varphi$ are polar angles in the spherical coordinate system; 
$r \in \{ - \infty, + \infty \}$
is the longitudinal coordinate; $Q$ is the magnetic charge, 
$r_0$ is the cross sectional size of the flux tube. This solution is a tube 
with the cross section $4 \pi r_0^2$, the coordinate $r$ is directed 
along the tube and the metric \eqref{sec1-10} is the 4D generalization 
of well known 4D Levi-Civita -- Bertotti -- Robinson solution 
\cite{levi}.
\par
On the 4D language we have the following components 
of the 4D electromagnetic potential
$A_\mu$
\begin{equation}
A_t =\sqrt{2} \sinh \left( \frac{r}{r_0} \right) \quad
\text{and} \quad
A_\varphi = Q \cos \theta = \sqrt{2} r_0 \cos \theta
\label{sec1-20}
\end{equation}
and the Maxwell tensor is 
\begin{equation}
    F_{rt} = \frac{\sqrt{2}}{r_0} \cosh \left( \frac{r}{r_0} \right) \quad
    \text{and} \quad
    F_{\theta \varphi} = - Q \sin \theta = - \sqrt{2} r_0 \sin \theta .
\label{sec1-30}
\end{equation}
The $(\chi t)$-Einstein's equation can be written in the following way :
\begin{equation}
    \left( 
    	4 \pi r_0^2 \frac{1}{\cosh \left( \frac{r}{r_0} \right)} 
    	\omega '  
    \right)' = 0.
\label{sec1-90}
\end{equation}
The 5D Kaluza - Klein gravity after the dimensional reduction indicates that
the Maxwell tensor is
\begin{equation}
    F_{\mu \nu} = \partial_\mu A_\nu - \partial _\nu A_\mu .
\label{sec1-100}
\end{equation}
It allows us to write the electric field as $E_r = \omega '$.
Eq.\eqref{sec1-90} with the electric field defined by \eqref{sec1-100}
can be compared with the Maxwell's equations in a continuous medium
\begin{equation}
    \mathrm{div} \mathcal {\vec D} = 0
\label{sec1-110}
\end{equation}
where $\mathcal {\vec D} = \epsilon \vec E$ is an electric displacement
and $\epsilon$ is a dielectric permeability. Comparing Eq. \eqref{sec1-110}
with Eq. \eqref{sec1-90} we see that the magnitude
$\frac{1}{\cosh \left( \frac{r}{r_0} \right)} \omega '$ 
is like to the electric displacement
and the dielectric permeability is 
$\epsilon = \frac{1}{\cosh \left( \frac{r}{r_0} \right)}$.
This means that $q = \sqrt{2} r_0$ can be taken 
as the Kaluza-Klein electric charge
because the flux of the electric field is
$\mathbf{\Phi} = 4\pi r_0^2 \mathcal D = 4\pi q = \text{const}$.
\par 
We see that this solution is a gravitational flux tube with 
the following properties :
\begin{itemize}
	\item the electric and magnetic charges are equal  
	\begin{equation}
		q = Q ;
	\label{sec1-40}
	\end{equation}	
	\item the radial electric displacement is equal to the radial 
	magnetic field 	
	\begin{equation}
		\mathcal D_r = H_r = 
		\frac{\sqrt{2}}{r_0} = \text{const};
	\label{sec1-50}
	\end{equation}
	\item the tube is filled with the constant parallel electric and 			  magnetic fields; 
	\item the linear size of the 5$^{th}$ dimension is $r_0$;  
	\item the cross sectional size of the flux tube is $r_0$ and can be 
	arbitrary. 
\end{itemize}
The last item allows us to consider the case $r_0 \rightarrow 0$ but with one essential  remark. If we believe that we live in Nature with a minimal (Planck) length then the  minimal value of $r_0$ is $\left( r_0 \right)_{min} = l_{Pl}$ and for a macroscopical  observer this minimal length is a point. It means that the presented solution in the limit  $\left( r_0 \right)_{min} = l_{Pl}$ is the flux tube with alomost zero cross section, i.e. a string-like object with a flux of electric and magnetic fields. Though the tube has some thickness  it is the  minimal one and for a macroscopical observer it is the string-like object. 
\par 
In the next section we will show that the solution presented here after $5D \rightarrow 4D$ reduction can be isometrically  embedded in a Minkowski spacetime and by $\left( r_0 \right)_{min} = l_{Pl}$ we will have a string-like object moving in an embedding spacetime (in a bulk). 

\section{Isometrical embedding of the flux tube}
\label{sec:IsometricalEmbedding}

Let us choose the Minkowski spacetime with the metric 
\begin{equation}
	ds^2 = \left( dt^1 \right)^2 + \left( dt^2 \right)^2 - 
	\left( dx^1 \right)^2 - \left( dx^2 \right)^2 - 
	\left( dx^3 \right)^2 - \left( dx^4 \right)^2.
\label{sec3-10}
\end{equation}
According to Kaluza - Klein point of view on the 5D gravity the 5D metric 
\eqref{sec1-10} can be considered as 4D metric 
\begin{equation}
	ds^2 = \cosh^2 \left( \frac{r}{r_0} \right) dt^2 - 
		dr^{2} - r^2_0 \left( d\theta ^{2} +
	\sin ^{2}\theta  d\varphi ^2 \right ) 
\label{sec3-20}
\end{equation}
and 4D electromagnetic field with the potential \eqref{sec1-20} 
and electric and magnetic fields \eqref{sec1-30}. From this point of view we will consider embedding of 4D metric \eqref{sec3-20} in 6D Minkowski 
spacetime \eqref{sec3-10}. One can see that this embedding can be realized by 
the following way 
\begin{align}
	t^1 & = r_0 \cosh\left(\frac{r}{r_0}\right) 
	\cos \left(\frac{t}{r_0}\right), 
	&t^2 &= r_0 \cosh\left(\frac{r}{r_0}\right) 
	\sin \left(\frac{t}{r_0}\right), 
	&x^1 & = r_0 \sinh\left(\frac{r}{r_0}\right) , 
\label{sec3-30}\\	
	x^2 &= r_0 \sin \theta \cos \phi, 
	&x^3 & = r_0 \sin \theta \sin \phi, 
	&x^4 &= r_0 \cos \phi .
\label{sec3-40}
\end{align}
Now we would like to show that this flux tube in the limit $r_0 \rightarrow 0$ can be  considered as a string-like object (gravitational flux tube with almost zero cross section). We see that the map of the 4D spacetime with the metric  \eqref{sec3-20} into a 6D Minkowski spacetime $\left( t^1, t^2, x^1, x^2, x^3, x^4 \right)$ can be factorized on 2D sphere in $\left ( x^2, x^3, x^4 \right )$ space with the equation 
\begin{equation}
	\left( x^2 \right)^2 + \left( x^3 \right)^2 + 
	\left( x^4 \right)^2 = r_0^2 
\label{sec3-45}
\end{equation}
and the hypersurface in the Minkowski spacetime 
$\left ( t^1, t^2, x^1 \right )$ with the equation 
\begin{equation}
	\left( t^1 \right)^2 + \left( t^2 \right)^2 = 
	r_0^2 + \left( x^1 \right)^2 .
\label{sec3-50}
\end{equation}
Let us consider the case $r_0 \rightarrow 0$. We can not put $r_0$ simply equal to zero as  in Nature exists a minimal length $l_{Pl}$. We can not say anything about the spacetime  
structure inside Planck region consequently the Planck volume is a physical point for the a  macroscopical observer. After this remark we can put $\left( r_0 \right)_{min} = l_{Pl}$.  In this case the 2D sphere in $\left ( x^2, x^3, x^4 \right )$ space becomes a point and 
equation \eqref{sec3-50} in the Minkowski spacetime describes the world sheet of a moving  string-like object. It is easy to see that any spacelike section of the worldsheet \eqref{sec3-50} is the moving string-like object. 

\section{The properties of the derived string-like object}
\label{sec:ThePropertiesOfTheDerivedString}

The most significant difference between derived string-like object and the ordinary string in string theory is that the derived here string moves in the spacetime with 
\textit{two times}. It results in that each point of the string moves on a closed path. 
It is very interesting that an observer living on a brane will see a moving particle but  not a string ! For example, if we consider the brane $t^1=$const then on the spacetime  spanned on $\left( t^1,x^1 \right)$ coordinates the string-like object is the world line of a tachyon.  Of curse one can choose another section of the worldsheet in such a way that it will be a  particle moving with the speed $v < c$. 
\par 
The equations of motion of the isometrically embedding string-like object in the Minkowski  spacetime are 5D Einstein's vacuum equations + the embedding conditions 
\begin{eqnarray}
	R_{(A)(B)} - \frac{1}{2} \eta_{(A)(B)} R &=& 0, 
\label{sec4-10}\\
	\left( \partial_A X^\mu \right) 
	\left( \partial_B X^\nu \right) \eta_{\mu \nu} &=& 
	G_{AB} = 
	e^{(A)}_A e^{(B)}_B \eta_{(A)(B)}
\label{sec4-20}
\end{eqnarray}
where $A,B = 0,1,2,3,5$ are 5 world indexes; $(C),(D)$ are 5-bein indexes; $X^\mu$ are the  coordinates of the embedded spacetime in the Minkowski spacetime \eqref{sec3-10}; 
$\eta_{(A)(B)}$ is the 5-bein metric; $\eta_{\mu \nu}$ is the metric of the 6D embedding Minkowski spacetime. 
\par 
The equations of the moving bosonic string in string theory are 
\begin{equation}
	\partial_A \partial^A X^\mu = 0 .
\label{sec4-30}
\end{equation}
We see that equations of motions \eqref{sec4-10}-\eqref{sec4-20} and \eqref{sec4-30}  are different. The reason is that the isometrically embedded gravitaional flux tube has a finite thickness.  The situation is similar to QCD string where the finite thickness of  QCD flux tube  leads to a rigidity and consequently the string action (and string equations) changes a little \cite{polyakov} \cite{kleinert}. 
\par 
There is an interesting difference between isometrical embedded gravitational string and  QCD string. In both cases the string has a finite thickness but in the first case the  electric and magnetic fields are concentrated inside flux tube only. In the second case the  color electric field is distributed in the whole space (though the most part of force  lines are concentrated inside the string).

\section{Gravitational waves on the string-like object}
\label{sec:GravitationalWavesOnTheString}

In this section we would like to show that on the gravitational flux tube \eqref{sec1-10} may exist some perturbations which after $r_0 \rightarrow 0$ or more exactly 
$\left( r_0 \right)_{min} = l_{Pl}$ will give us some perturbation of the worldsheet. 
\par 
Let us introduce the 5-bein for the perturbed metric 
\begin{eqnarray}
	e^{(A)}_B e_{(A)C} &=& G_{BC}, 
\label{sec5-10}\\
	e_{(A) C} e_{(B)}^C &=& \eta_{(A) (B)} = 
	\text{diag}\left\{ +1,-1,-1,-1,-1 \right\}
\label{sec5-20}
\end{eqnarray}
where $G_{AB}$ is the perturbed metric \eqref{sec1-10}. 
We will consider the next perturbation 
\begin{eqnarray}
	e^{(A)}_B &=& \stackrel{0}{e}^{(A)}_B + 
	\widetilde{e}^{(A)}_B, 
\label{sec5-30}\\
	e^{(0)}_0 &=& \cosh \left(\frac{r}{r_0} \right) + 
	\widetilde{e}^{(0)}_0, \quad 
	e^{(1)}_1 = 1 + \widetilde{e}^{(1)}_1
\label{sec5-40}
\end{eqnarray}
where $\stackrel{0}{e}^{(A)}_B$ is the 5-bein for the metric \eqref{sec1-10}; $\widetilde{e}^{(A)}_B$ are the perturbations. 
The perturbed metric is 
\begin{equation}
\begin{split}
	ds^2 & = \left[ 
		\cosh \left( \frac{r}{r_0} \right) + \widetilde{e}^{(0)}_0
	\right]^2 dt^2 - 
	\left[
		1 + \widetilde{e}^{(1)}_1
	\right]^2	dr^{2} - r^2_0 \left( d\theta ^{2} +
	\sin ^{2}\theta  d\varphi ^2 \right ) 
\\
	& -	\left [d\chi + 
	\omega(r) dt + 
	Q \cos \theta d\varphi \right ]^2 , 
\\ 
	\omega(r) & = \sqrt{2} \sinh \left( \frac{r}{r_0} \right),
	\quad 
	Q = \sqrt{2} r_0
\label{sec5-45}
\end{split}
\end{equation}
Let us note that we consider the perturbation of the 4D metric only, 
the perturbations of $G_{55}$ and the electromagnetic field 
$A_\mu = \widetilde{e}^{(5)}_\mu$ are frozen. Additionally we take 
\begin{equation}
	\widetilde{e}^{(0)}_0 = - \widetilde{e}^{(1)}_1 
	\cosh \left(\frac{r}{r_0} \right) .
\label{sec5-50}
\end{equation}
Then the perturbed 5D Einstein's equations give us 
\begin{equation}
	\frac{\partial^2 \widetilde{e}^{(1)}_1}{\partial \tau^2} + 
	\cosh^2 x \frac{\partial^2 \widetilde{e}^{(1)}_1}{\partial x^2} + 
	3 \sinh x \cosh x \frac{\partial \widetilde{e}^{(1)}_1}{\partial x} + 
	2 \cosh^2 x \widetilde{e}^{(1)}_1 = 0 
\label{sec5-60}
\end{equation}
where $\tau = t/r_0$ and $x = r/r_0$ are correspondingly 
the dimensionless time and longitudinal coordinates. 
\par 
One can search the solution in the form 
\begin{equation}
	 \widetilde{e}^{(1)}_1 = T(t) X (x)
\label{sec5-70}
\end{equation}
with the following equations for $T(t)$ and $X (x)$ 
\begin{eqnarray}
	 \ddot{T}(t) &=& -\omega^2 T(t),
\label{sec5-80}\\
	X'' + 3 \tanh x \; X' + 
	\left( 2 - \frac{\omega^2}{\cosh^2 x} \right)X &=& 0
\label{sec5-90}
\end{eqnarray}
Eq's \eqref{sec5-80} \eqref{sec5-90} have the following special solutions 
\begin{alignat}{3}
	 T(t) &= A \sin \left( \omega \tau + \alpha\right),
	 &\quad X(x) &= \frac{C}{\cosh x} \ \ 
	 &\quad \text{for } \omega^2 &= +1,
\label{sec5-100}\\
	 T(t) &= D e^{ \left| \omega \right| \tau} + 
	 E e^{- \left| \omega \right| \tau},
	 &X(x) &= F \frac{\sinh x}{\cosh^3 x} \ \ 
	 &\text{for } \omega^2 &= -3,
\label{sec5-110}
\end{alignat}
i.e. oscillating gravitational waves 
\begin{equation}
	 \widetilde{e}^{(1)}_1 = G 
	 \frac{\sin ( \tau + \alpha)}{\cosh x}, 
	 \widetilde{e}^{(0)}_0 = - G 
	 \sin ( \tau + \alpha) 
\label{sec5-130}
\end{equation}
and damped gravitational waves 
\begin{equation}
	 \widetilde{e}^{(1)}_1 = H 
	 \frac{\sinh x}{\cosh^3 x} e^{-\sqrt 3 \: \tau}, 
	 \widetilde{e}^{(0)}_0 = - H 
	 \frac{\sinh x}{\cosh^2 x} e^{-\sqrt 3 \: \tau} 
\label{sec5-140}
\end{equation}
where $A, C, D, F, G, H, \alpha$ are some constants. 
The exponential increasing gravitational waves with 
$T(\tau) = D e^{\sqrt 3 \tau}$ have to be investigated more 
carefully as for the big time $\tau$ such perturbation analysis 
is incorrect. 
\par 
Evidently the perturbed 4D metric 
\begin{equation}
	ds^2 = \left[ 
		\cosh \left( \frac{r}{r_0} \right) + \widetilde{e}^{(0)}_0
	\right]^2 dt^2 - 
	\left[
		1 + \widetilde{e}^{(1)}_1
	\right]^2	dr^{2} - r^2_0 \left( d\theta ^{2} +
	\sin ^{2}\theta  d\varphi ^2 \right ) 
\label{sec5-150}
\end{equation}
can be embedded in some multidimensional Minkowski spacetime with more than two times.

\section{Discussion and conclusions}
\label{sec:DiscussionAndConclusions}

In this notice we have shown that in 5D Kaluza-Klein gravity exists solution which after 
$5D \rightarrow 4D$ reduction can be isometrically embedded in the 6D Minkowski spacetime 
with two times. We have considered here the case with the infinite gravitational flux tube  $(q = Q)$. In Ref. \cite{dzhsin2} it is shown that there is the case with a finite gravitational flux tube solution $\left( q \approx Q, q > Q \right)$. Such finite flux tube with the cross section in the Planck region can be isometrically embedded in a Minkowski spacetime as well. The difference  with the infinite flux tube solution $(q = Q)$ is that in the case $ \left(q \approx Q, q > Q \right)$ the whole spacetime looks as two spacetimes 
connected by a very long gravitational flux tube $\left( 0 \leq 1 - \frac{q}{Q} \ll 1 \right)$ filled  with almost equal longitudinal electric and magnetic fields. After isometrical embedding  into a Minkowski spacetime with several times this solution will look as two branes connected with a string-like object. 
\par 
The investigation presented here shows that if the cross section of such flux tube is in Planck region then such tube can be considered as a string-like object but with some differences with the ordinary string: the string-like object has finite thickness, moves in 6D Minkowski spacetime with two times and equations of motions for string and string-like object are different. 

\section{Acknowledgments}

I am very grateful D. Singleton for the fruitful comments.

\end{document}